\documentstyle[preprint,aps]{revtex}
\tighten
\draft
\begin{document}
\widetext
\preprint{CERN-TH/97-155, HUTP-97/A033, NUB 3163}
\bigskip
\bigskip
\title{Dynamical Flavor Hierarchy and Heavy Top}
\medskip
\author{Gia Dvali$^{1}$\footnote{E-mail: 
georgi.dvali@cern.ch} and 
Zurab Kakushadze$^{2,3}$\footnote{E-mail: 
zurab@string.harvard.edu}}

\bigskip
\address{$^1$Theory Division, CERN, CH-1211, Geneva 22, Switzerland\\
$^2$Lyman Laboratory of Physics, Harvard University, Cambridge, 
MA 02138\\
$^3$Department of Physics, Northeastern University, Boston, MA 02115}
\date{\today}
\bigskip
\medskip
\maketitle

\begin{abstract}

{}In the Standard Model one of the generations is much heavier than the other two. We propose a simple mechanism where all three generations are treated on the equal footing to begin with, and one heavy and two light families 
emerge from supersymmetric strong dynamics. The Yukawa mass matrix is identified with vevs of the meson fields of some additional gauge theory. It is then forced to have rank one (in the leading order) by non-perturbative superpotential.

\end{abstract}
\narrowtext

{}The origin of flavor hierarchy has been a puzzle ever since the Standard Model of strong and electroweak interactions emerged. Recently \cite{new}, as well as in the past \cite{old}, much effort has gone into building models where such hierarchy can occur as a result of non-Abelian flavor symmetry. (Such symmetries
have appeared in recently derived three-family grand unified string models \cite{kt}.) Ideally one would like to obtain the observed flavor hierarchy dynamically. One of the most appealing scenarios would be if this dynamics is that of a ($N=1$ supersymmetric) strongly coupled non-Abelian gauge theory. With the recent progress \cite{rev} in understanding many of the non-perturbative aspects of such theories, including exact superpotentials, it becomes even more desirable to look into a possibility of solving the flavor hierarchy problem along these lines.

{}There are two ways of looking at the flavor hierarchy puzzle. First, one can ask why the first two generations are so much lighter than the third one. But one can turn the question around and ask whether there is a good reason for one of the generations to be much heavier than the other two. Most of the time the matter is considered from the first point of view. Then one tries to dynamically generate small numbers to explain the ratios of the first and second generation quark and lepton masses to those of the third generation. The drawback here is that the top-quark being heavy is taken for granted, hence inequality in treating different generations to begin with. Recently, some models along these lines have been constructed in Ref \cite{Hall}. In all of these models, however,
there is no symmetry between the third and the other two generations from the start, and, therefore, heavy top is an input rather than a prediction of the models. In this letter we address the issue from the point of view where one would like to {\em a priori} treat all three generations on the equal footing and generate one heavy and two light generations dynamically. In the following we propose a simple mechanism, and a model where the latter is realized, which dynamically generates large mass for one of the generations, while leaving the other two light.  

{}First, let us formulate the problem. The Standard Model gauge group is $SU(3)_c \otimes SU(2)_w \otimes U(1)_Y$. Let $Q_i$ be the left-handed quarks transforming in $({\bf 3},{\bf 2})$ of $SU(3)_c \otimes SU(2)_w$. Let ${\tilde Q}^{\bar j}$ be the left-handed anti-quarks transforming in $({\overline {\bf 3}},{\bf 1})$ of $SU(3)_c \otimes SU(2)_w$. There is an $SU(3)_L\otimes SU(3)_R$ global flavor symmetry under which $Q_i$ transform as $({\overline {\bf 3}},{\bf 1})$, while 
${\tilde Q}^{\bar j}$ transform as $({\bf 1},{\bf 3})$. Then $i$ and ${\bar j}$ are the corresponding flavor indices. Next, consider the following Yukawa couplings: ${\cal Y}^i_{\bar j} Q_i {\tilde Q}^{\bar j} H$. Here ${\cal Y}^i_{\bar j}$ is the Yukawa mass matrix, and $H$ is the corresponding electroweak Higgs doublet. The question is whether we can find a dynamical mechanism which would force ${\cal Y}^i_{\bar j}$ to have rank one (in the first approximation), and at the same time naturally make the only non-zero entry of   
${\cal Y}^i_{\bar j}$ (in the diagonal basis) have value of the order of one.

{}The idea we propose is very simple. First, we would like to view ${\cal Y}^i_{\bar j}$ as vevs of some holomorphic field ${ { Y}}^i_{\bar j}$: ${\cal Y}^i_{\bar j}=\langle { { Y}}^i_{\bar j} \rangle$. This is much in the spirit of Seiberg's holomorphicity principle. Next, (up to a dimensionful constant-see below) we identify ${ { Y}}^i_{\bar j}$ with a low energy {\em meson} field ${M}^i_{\bar j}$ of some additional strongly coupled non-Abelian gauge theory. That is, we identify the global flavor symmetry group of this strongly coupled theory with that of the Standard Model, {\em i.e.}, $SU(3)_L\otimes SU(3)_R$ global flavor group of the quarks $Q_i$ and ${\tilde Q}^{\bar j}$. The constraint on ${\cal Y}^i_{\bar j}\sim {M}^i_{\bar j}$ that guarantees the correct (first approximation) flavor hierarchy is then supposed to emerge from the strong coupling dynamics of the additional non-Abelian gauge theory.   

{}Next, we give a simple model that puts flesh on the idea we just described. On top of the Standard Model gauge group $SU(3)_c \otimes SU(2)_w \otimes U(1)_Y$ we introduce $SU(3)_s$ gauge symmetry with some dynamical scale $\Lambda$ that lies above the supersymmetry breaking scale $M_{SUSY}$ in the observable sector.
We have the usual quarks $Q_i$ and ${\tilde Q}^{\bar j}$, Higgs doublet $H$, and also the quarks $q^i$ and ${\tilde q}_{\bar j}$ of $SU(3)_s$. The massless spectrum of this model is summarized in Table I.

{}Now, let us assume that there is a tree-level superpotential
\begin{equation}
 {\cal W}_{tree}=\lambda B +{\tilde \lambda} {\tilde B}+{1\over {M^2_{Pl}}}
 {M}^i_{\bar j} Q_i {\tilde Q}^{\bar j} H~.
\end{equation}
Here 
\begin{equation}
 {M}^i_{\bar j}\equiv q^i {\tilde q}_{\bar j}~,~~~B\equiv \epsilon_{ijk}
 q^i q^j q^k ~,~~~
 {\tilde B}\equiv \epsilon^{{\bar i}{\bar j}{\bar k}} 
 {\tilde q}_{\bar i} {\tilde q}_{\bar j} {\tilde q}_{\bar k}
\end{equation}
are the low energy mesons and baryons ($SU(3)_s$ singlets), whereas $\lambda$ and ${\tilde \lambda}$ are three-level Yukawa couplings whose natural values are of the order of one. (Note that $\lambda B$ and ${\tilde \lambda}{\tilde B}$ are three-point couplings in terms of the fundamental quarks $q^i$ and anti-quarks ${\tilde q}_{\bar j}$. The term ${M}^i_{\bar j} Q_i {\tilde Q}^{\bar j} H/M^2_{Pl}$ is a five-point non-renormalizable coupling once written in terms of the fundamental quarks and anti-quarks. We will comment on possible origins of this term at the end of this paper. Note, however, that such non-renormalizable
couplings are inevitably present in all models with non-Abelian flavor symmetries.)

{}Note that as far as the $SU(3)_s$ theory is concerned it is an $N=1$ supersymmetric $SU(N_c)$ theory with $N_f=N_c$ flavors ($N_c=3$ in our case). This theory has quantumly modified moduli space and the non-perturbative superpotential that is generated reads \cite{Seiberg}: 
\begin{equation}
 {\cal W}_{non-pert}=A(\det(M)-B{\tilde B}-\Lambda^6)~.
\end{equation}
Here $A$ is a Lagrange multiplier. The exact superpotential then is given by
\begin{equation}
 {\cal W}={\cal W}_{tree}+{\cal W}_{non-pert}=
 A(\det(M)-B{\tilde B}-\Lambda^6)+\lambda B +{\tilde \lambda} {\tilde B}+{1\over   
 {M^2_{Pl}}}{M}^i_{\bar j} Q_i {\tilde Q}^{\bar j} H~.
\end{equation}

{}Let us study the flat directions in this superpotential. The $F$-flatness 
conditions read:
\begin{eqnarray}\label{Fflat}
 &&0=F_A={{\partial {\cal W}}\over {\partial {A}}}=
 \det(M)-B{\tilde B}-\Lambda^6 ~,\\
 &&0=F_B=
 {{\partial {\cal W}}\over {\partial B}}=-A{\tilde B}+\lambda~,\\
 &&0=F_{\tilde B}=
 {{\partial {\cal W}}\over {\partial {\tilde B}}}=-A{B}+{\tilde
 \lambda}~,\\
 &&0=F_{M^i_{\bar j}}={{\partial {\cal W}}\over {\partial M^i_{\bar j}}}=
 A{\cal M}_i^{\bar j}
 +{1\over {M^2_{Pl}}} Q_i {\tilde Q}^{\bar j} H~,
\end{eqnarray}
where
\begin{equation}
 {\cal M}_i^{\bar j}\equiv{{\partial\det(M)}\over {\partial M^i_{\bar j}}}~. 
\end{equation}
We assume that the vevs of the Standard Model quarks $Q_i$ and ${\tilde Q}^{\bar j}$ are zero (as they should be, or else $SU(3)_c$ would be broken). Then we have the following solution to the above $F$-flatness conditions (provided that $\lambda,{\tilde
\lambda}\not=0$):
\begin{equation}\label{F}
 A={\sqrt{-\lambda{\tilde{\lambda}}}\over{\Lambda^3}}~,~~~
 B=\Lambda^3 \sqrt{-{{\tilde{\lambda}\over \lambda}}}~,~~~
 {\tilde B}=\Lambda^3 \sqrt{-{ \lambda\over{\tilde{\lambda}}}}~,~~~ 
 {\cal M}_i^{\bar j}=0~.
\end{equation}
Note that since
\begin{equation}
  3\det(M)=M^i_{\bar j}{\cal M}_i^{\bar j}~,
\end{equation}
the last condition in Eq (\ref{F}) implies that $\det(M)=0$. Since all the $2\times 2$ minors ${\cal M}_i^{\bar j}$ of the matrix $M^i_{\bar j}$ are also zero, we see that the rank of the latter can be at most one. In the diagonal basis we can, therefore, set $M^i_{\bar j}={\mbox{diag}}(0,0,M^3_{\bar 3})$. Further, note that $M^3_{\bar 3}$ is a flat modulus so that upon supersymmetry breaking its stabilized value can be of the order of $M^2_{Pl}$. If this is the case, then the net result is that the Yukawa mass matrix ${\cal Y}^i_{\bar j}$ has rank one, and its only non-zero element (in the diagonal basis) is of the order of one. Thus, we have one heavy generation (corresponding to the-top quark), and two light 
generations. Note that this flavor hierarchy arises in this model dynamically, and never did we treat any of the generations on a different footing than others. Also note that the scale $\Lambda$ for the above matter content of $SU(3)_s$ is always higher than the supersymmetry breaking scale in the observable sector. 

{}Here a few comments are in order. First note that the above solution to the
$F$-flatness conditions is only valid if both $\lambda$ and ${\tilde \lambda}$ 
are non-zero. If both of them are zero, then we do {\em not} get a restriction
on $\det(M)$ as the Lagrange multiplier $A$ can take zero value. If only one 
of them is zero, say, ${\tilde \lambda}$, then there is a possibility of having
dynamical breaking of {\em local} supersymmetry on top of the dynamically generated flavor hierarchy we just described. Indeed, in this case the $F$-flatness conditions (\ref{Fflat})
cannot be simultaneously satisfied for any finite values of the $A$, $B$, ${\tilde B}$ and $M^i_{\bar j}$ vevs (provided that $Q_i$ and $Q^{\bar j}$ do not acquire vevs), {\em i.e.}, there is a runaway direction, namely, ${\tilde B}$, along which the superpotential is non-zero for any finite ${\tilde B}$.
It can be shown that if in a globally supersymmetric theory for any finite values of vevs $F$-flatness conditions cannot be satisfied and the superpotential does not vanish, then in the locally supersymmetric version of this theory supersymmetry will be broken due to the K{\"a}hler potential contributions. Intuitively this can be understood by noting that once such a theory is coupled to supergravity there is a natural shut-down scale for all the runaway directions, namely, the Plank scale. Generically, this results in local supersymmetry breakdown. Thus, local supersymmetry may be broken in the above model with ${\tilde \lambda}=0$.

{}Let us return to the flavor hierarchy problem. We see that in the above simple model due to the non-perturbative dynamics in the $SU(3)_s$ sector one of the generations comes out heavy, while the other two are light. Naturally, one wonders if the more fine structure in the flavor hierarchy, such as, say, light quark/lepton masses and mixing angles, can also be obtained in this way. Here we point out one possible source for such fine structure. There can be higher dimensional operators (suppressed by $M_{Pl}$), some of which can explicitly break the $SU(3)_L\otimes SU(3)_R$ flavor symmetry. For instance, one can couple higher $SU(3)_s$ invariants to the Standard Model quarks in the spirit of Ref \cite{Hall}.

{}Finally, we would like to comment on the origin of the non-renormalizable (five-point) term ${M}^i_{\bar j} Q_i {\tilde Q}^{\bar j} H/M^2_{Pl}$ in the tree-level superpotential. Certainly, it could naturally arise in string theory. It could either directly be present in a particular string model, or emerge from three-point renormalizable Yukawa couplings once, say, some additional vector-like states are integrated out. (The latter type of scenarios were considered in the field theory context in Ref \cite{zura}. In certain cases integrating out such fields may result in dangerous flavor non-universal contributions to the squark masses \cite{alex}.) We leave this to the reader's taste.

\acknowledgments

{}We would like to thank Alex Pomarol and Riccardo Rattazzi for discussions.
The work of Z.K. was supported in part by the grant NSF PHY-96-02074, and the DOE 1994 OJI award. Z.K. would like to thank CERN Theory Division for their kind hospitality while this work was completed. Z.K. would also like to thank Albert and Ribena Yu for financial support.

\begin{table}[t]
\begin{tabular}{|c|c|c|c|c|c|}
&$SU(3)_c$&$SU(2)_w$&$SU(3)_s$&$SU(3)_L$&$SU(3)_R$\\
\hline
$Q_i$&${\bf 3}$&${\bf 2}$&${\bf 1}$&${\overline {\bf 3}}$&${\bf 1}$\\
${\tilde Q}^{\bar j}$&${\overline{\bf 3}}$&${\bf 1}$&${\bf 1}$&
 ${\bf 1}$&${\bf 3}$\\
$H$&${\bf 1}$&${\bf 2}$&${\bf 1}$&${\bf 1}$&${\bf 1}$\\
$q^i$&${\bf 1}$&${\bf 1}$&${\bf 3}$&${\bf 3}$&${\bf 1}$\\
${\tilde q}_{\bar j}$&${\bf 1}$&${\bf 1}$&${\overline{\bf 3}}$&
${\bf 1}$&${\overline{\bf 3}}$\\
\end{tabular}
\caption{The local $SU(3)_c\otimes SU(2)_w\otimes SU(3)_s$ and global 
$SU(3)_L\otimes SU(3)_R$ charges of the matter fields in the model described in this paper.}
\end{table}


\begin{references}

\bibitem{new} See, {\em e.g.},\\
M. Dine, R. Leigh and A. Kagan, Phys. Rev. {\bf D48} (1993) 4269;\\
Y. Nir and N. Seiberg, Phys. Lett. {\bf B309} (1993) 337;\\
P. Pouliot and N. Seiberg, Phys. Lett. {\bf B318} (1993) 169;\\
D. Kaplan and M. Schmaltz, Phys. Rev. {\bf D48} (1993) 4269;\\
R. Barbieri, G. Dvali and A. Strumia, Nucl. Phys. {\bf B435} (1995) 102;\\
A. Pomarol and D. Tommasini, Nucl. Phys. {\bf B466} (1996) 3;\\
L.J. Hall and H. Murayama, Phys. Rev. Lett. {\bf 75} (1995) 3985;\\
P.H. Frampton and O.C.W. Kong, Phys. Rev. {\bf D53} (1995) 2293;
Phys. Rev. Lett. {\bf 77} (1996) 1699;\\
P.H. Frampton and T.W. Kephart, Phys. Rev. {\bf D51} (1995) 1;\\
R. Barbieri, G. Dvali and L.J. Hall, Phys. Lett. {\bf B337} (1996) 76;\\
N. Arkani-Hamed, H.C. Cheng and L.J. Hall, 
Nucl. Phys. {\bf B472} (1996) 95; Phys. Rev. {\bf D54} (1996) 2242;\\
K.S. Babu and S.M. Barr, Phys. Lett. {\bf B387} (1996) 87;\\
R. Barbieri and L.J. Hall, Nuovo Cim. {\bf 110A} (1997) 1;\\
Z. Berezhiani, Nucl. Phys. Proc. Suppl. {\bf 52A} (1997) 153; hep-ph/9609342;\\
A. Ra{\v s}in, hep-ph/9705210.

\bibitem{old} For some early works, see, {\em e.g.},\\
F. Wilczek and A. Zee, Phys. Rev. Lett. {\bf 42} (1979) 421;\\
J. Chkareuli, Pis'ma Zh. Eksp. Teor. Fiz. {\bf 32} (1980) 684 [JETP Lett. {\bf 32} (1980) 671];\\
Z. Berezhiani and J. Chkareuli, Yad. Fiz. {\bf 37} (1983) 1043 [Sov. J. Nucl. Phys. {\bf 37} (1983) 618].

\bibitem{kt} See, {\em e.g.},\\
Z. Kakushadze and S.-H.H. Tye, Phys. Rev. Lett. {\bf 77} 
(1996) 2612; Phys. Rev. {\bf D54} (1996) 7520; 
Phys. Rev. {\bf D55} (1997) 7878; Phys. Rev. {\bf D55} (1997) 7896.

\bibitem{rev} See, {\em e.g.},\\
N. Seiberg, Nucl. Phys. {\bf B435} (1995) 129;\\
K. Intriligator and N. Seiberg, Nucl. Phys. {\bf B444} (1995) 125; Nucl. Phys. Proc. Suppl. {\bf 45BC} (1996) 1.

\bibitem{Hall} C.D. Carone, L.J. Hall and T. Moroi, hep-ph/9705383.

\bibitem{Seiberg} N. Seiberg, Phys. Rev. {\bf D49} (1994) 6857.

\bibitem{zura} C.D. Froggatt and H.B. Nielsen, Nucl. Phys. {\bf B147} (1979) 277;\\
Z. Berezhiani, Phys. Lett. {\bf B129} (1983) 99; Phys. Lett. {\bf B150} (1985) 177;\\
S. Dimopoulos, Phys. Lett. {\bf B129} (1983) 417. 

\bibitem{alex} A. Pomarol and D. Tommasini (Ref \cite{new}).

\end{references}
\end{document}